\documentclass[manuscript]{acmart}

\usepackage[utf8]{inputenc}
\usepackage{tabulary}
\usepackage{booktabs}
\usepackage{multirow}
\usepackage{siunitx}
\usepackage{pifont}
\usepackage{xcolor}

\newcommand{\cmark}{\ding{51}}%
\newcommand{\xmark}{\ding{55}}%

\newif\ifworkinprogress
\workinprogresstrue

\ifworkinprogress
	\newcommand{\ap}[1]{\textcolor{green}{{[Andreas] #1}}}
	\newcommand{\ez}[1]{\textcolor{red}{{[Eva] #1}}}

\else
  \newcommand{\ap}[1]{}
  \newcommand{\ez}[1]{}
\fi

\newcommand{\eg}{e.\,g., }
\newcommand{\ie}{i.\,e., }

\newcommand{\Ni}{(1)~}
\newcommand{\Nii}{(2)~}
\newcommand{\Niii}{(3)~}

\AtBeginDocument{%
  \providecommand\BibTeX{{%
    \normalfont B\kern-0.5em{\scshape i\kern-0.25em b}\kern-0.8em\TeX}}}

\title[Unsupervised Graph Embeddings for Session-based Recommendation]{Unsupervised Graph Embeddings for Session-based Recommendation with Item Features}

\author{Andreas Peintner}
\affiliation{%
  \institution{University of Innsbruck}
  \city{Innsbruck}
  \country{Austria}
  }
\email{andreas.peintner@uibk.ac.at}

\author{Marta Moscati}
\affiliation{%
  \institution{Johannes Kepler University Linz}
  \city{Linz}
  \country{Austria}
  }
\email{marta.moscati@jku.at}

\author{Emilia Parada-Cabaleiro}
\affiliation{%
  \institution{Johannes Kepler University Linz and Linz Institute of Technology}
  \city{Linz}
  \country{Austria}
  }
\email{emilia.parada-cabaleiro@jku.at}

\author{Markus Schedl}
\affiliation{%
  \institution{Johannes Kepler University Linz and Linz Institute of Technology}
  \city{Linz}
  \country{Austria}
  }
\email{markus.schedl@jku.at}

\author{Eva Zangerle}
\affiliation{%
  \institution{University of Innsbruck}
  \city{Innsbruck}
  \country{Austria}
  }
\email{eva.zangerle@uibk.ac.at}

\setcopyright{rightsretained}
\acmYear{2022}
\acmConference[RecSys ’22, CARS Workshop]{}{September 23rd, 2022}{Seattle, WA, USA}
\acmBooktitle{CARS: Workshop on Context-Aware Recommender Systems (RecSys ’22), September 23rd, 2022, Seattle, WA, USA}

\begin{document}

\begin{abstract}
  In session-based recommender systems, predictions are based on the user's preceding behavior in the session. State-of-the-art sequential recommendation algorithms either use graph neural networks to model sessions in a graph or leverage the similarity of sessions by exploiting item features. In this paper, we combine these two approaches and propose a novel method, \emph{\textbf{G}raph \textbf{C}onvolutional \textbf{N}etwork \textbf{Ext}ension (GCNext)}, which incorporates item features directly into the graph representation via graph convolutional networks. 
  \textit{GCNext} creates a feature-rich item co-occurrence graph and learns the corresponding item embeddings in an unsupervised manner. 
  We show on three datasets that integrating GCNext into sequential recommendation algorithms significantly boosts the performance of nearest--neighbor methods as well as neural network models. Our flexible extension is easy to incorporate in state-of-the-art methods and increases the \emph{MRR@20} by up to $12.79\%$.
  
\end{abstract}

\maketitle

\section{Introduction}
Recommender systems (RecSys) traditionally leverage the users' rich interaction data with the system. However, in some cases, such data are not available. 
Session-based recommender systems, in contrast, aim to predict the next item the user will interact with (\eg click on, purchase, or listen to) only based on the preceding interactions in the current session. 
The task of session-based recommendation can be defined as follows. Consider the set $X$  of all items in the catalog, $x \in X$ being an individual item, and $m = |X|$ being the number of items in the catalog. 
Given an interaction session $[x_1,x_2,...,x_n]$ (ordered by timestamp), the goal is to predict a ranked list $[y_1, y_2,...,y_m]$ of items with corresponding relevance scores to continue the session. 
The top-$k$ values of the ranked list are chosen as recommendation candidates. As opposed to session-aware or sequential recommendations, the inputs to session-based RecSys are only items of the current session and their features; users are anonymous and no inter-session data is available.

Current approaches for session-based recommendation leverage Recurrent Neural Networks (RNNs)~\cite{hidasi2015session,tan2016improved,ren2019repeatnet}, attention networks~\cite{li2017neural,liu2018stamp}, Graph Neural Networks (GNNs)~\cite{pan2021dynamic,wu2019session,xu2019graph}, or transformer architectures~\cite{kang2018self,sun2019bert4rec,de2021transformers4rec}. Also, classical nearest--neighbor methods have been used
~\cite{ludewig2021empirical,ludewig2018effective,jannach2017recurrent,garg2019sequence}. 
%
Most current methods focus on the sequential nature of sessions; RNNs and nearest--neighbor methods have dominated research in the past few years. Extensions to these models use additional item features to enrich the item representations. Item features capture contextual information (\eg item category) which is relevant to the task of session-based recommendation, which itself can be considered a special case of context-aware recommender systems~\cite{seq2018quadrana}. However, recently, GNN models have been shown to outperform RNN- and nearest-neighbor--based methods~\cite{ludewig2021empirical,wu2019session,gwadabe2022ic}. Yet, to the best of our knowledge, no approach combines auxiliary item features and GNNs to learn informative embeddings for sequential models. 
In this paper, we therefore propose \emph{\textbf{G}raph \textbf{C}onvolutional \textbf{N}etwork \textbf{Ext}ension (GCNext)}, which extracts node embeddings from a feature-rich item co-occurrence graph via unsupervised learning with Graph Convolutional Networks (GCNs). We then use these pre-trained item embeddings as auxiliary features describing items and their structural dependence. One major advantage and novelty of GCNext is that it can flexibly be plugged into any current sequential recommendation method. Particularly, we \Ni use the computed item embeddings to initialize sequential neural network models, and \Nii extend (non-neural) nearest--neighborhood methods with pre-trained item graph embeddings to refine the search of candidate sessions for recommendation.

Our main contributions are as follows: 
\Ni We present GCNext, a novel method for session-based recommendation based on a item co-occurence graph for sessions. GCNext combines the topological representation power of GCNs with the session representation generated by neural network sequential models without modifying their architecture; 
\Nii GCNext can easily extend nearest-neighbor methods as well as neural network models in a plug-in fashion to further enhance the performance of these models; 
\Niii We perform a large-scale evaluation of graph-based item embeddings and their impact on a diverse set of sequential models. We find that adding GCNext is not only able to boost the performance of current methods, but also shows significant performance improvements over current state-of-the-art sequential models.
\section{Related Work}
In the following, we briefly present related works in the field of graph and node embedding. We subsequently discuss approaches for sequential and session-based recommendation, which incorporate side information or GNNs.


\subsection{Graph and Node Embeddings}
Graph embedding aims to generate low-dimensional vector representations of the graph's nodes which preserve topology and leverage node features. Non-deep learning methods are mainly based on random walks to explore node neighborhoods~\cite{perozzi2014deepwalk,tang2015line,grover2016node2vec}. With GCNs~\cite{kipf2016semi,velickovic2018graph}, more sophisticated graph embedding methods were introduced: To scale GCNs to large graphs, the layer sampling algorithm~\cite{hamilton2017inductive} generates embeddings from a fixed node neighborhood. Current state-of-the-art methods in unsupervised learning of representations rely on contrastive methods which base their loss on the difference between positive and negative samples. Deep Graph Infomax (DGI)~\cite{velickovic2019deep} contrasts node and graph encodings by maximizing the mutual information between them. Hassani and Khasahmadi~\cite{hassani2020contrastive} propose multi-view representation learning by contrasting first-order neighbor encodings with a general graph diffusion. 
Contrastive learning methods usually require a large number of negative examples and are, therefore, not scalable for large graphs. The approach by Thakoor et al.~\cite{thakoor2021bootstrapped} learns by predicting substitute augmentations of the input and circumventing the need of contrasting with negative samples.

\subsection{Sequential Recommendation}
Non-neural sequential recommendation approaches focus on the similarity of sessions to extract potential next items. Several works extend the session-based nearest-neighbors method with additional factors such as positions, recency, and popularity~\cite{ludewig2021empirical,ludewig2018effective,jannach2017recurrent,garg2019sequence}. Other works~\cite{tan2016improved,li2017neural,liu2018stamp,kang2018self} model item-to-item transitions using neural networks, possibly incorporating item features~\cite{hidasi2016parallel,zhang2019feature,de2021transformers4rec}.

Recent works exploit the graph-based representation of sessions for improved recommendations. Current state-of-the-art use GNNs---in combination with attention or self-attention modules---to capture complex transitions and rich local dependencies~\cite{wu2019session,xu2019graph}. Further approaches enrich the graph topology with knowledge base entities~\cite{wang2020knowledge,amjadi2021katrec}. Gwadabe and Liu~\cite{gwadabe2022ic} use an item co-occurrence graph to generate session co-occurrence representations which are combined with the local and global preferences of users.

In contrast to models that integrate item features by extending the network with additional paths, GCNext extracts item embeddings from the item co-occurrence graph, in which content-based features are attached to each node. GCNext can be added to different sequential models without modifying their architecture, essentially using it in a plug-in fashion. Compared to already existing graph-based pretraining schemes~\cite{pretrain2021meng,multimodal2022liu} for general recommendation, our approach specifically tackles the task of sequential and session-based recommendations. 

\section{Graph Convolutional Network Extension (GCNext)}
\begin{figure}[h]
  \caption{Overview of the graph-based generation of item embeddings and its application in sequential neural network models.}
  \centering
  \includegraphics[width=\linewidth]{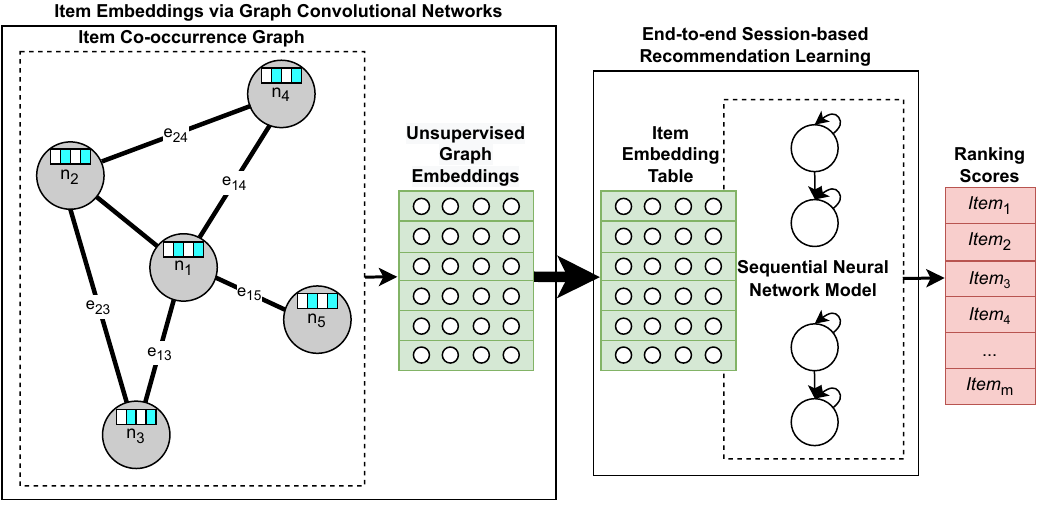}
  \label{fig:overview}
\end{figure}

In this section, we present the proposed GCNext approach. An overview 
of GCNext applied to sequential neural network models is shown in Figure~\ref{fig:overview}. The first component represents the item co-occurrence graph from which we extract corresponding node embeddings by applying an unsupervised learning method. We subsequently use these embeddings to initialize the item embedding table of the underlying end-to-end sequential model, which learns session-based recommendations. Furthermore, we show how GCNext can also be employed for nearest--neighbor methods.

\subsection{Unsupervised Graph Embeddings}


In contrast to other graph-based models~\cite{wu2019session,xu2019graph}, we generate the item-item graph by extracting item co-occurrences in sessions. Each item is modeled as a node with the item's features as descriptors. For two nodes $n_{1}$ and $n_{2}$, an edge between the nodes represents the co-occurrence of two items $i_1$ and $i_2$ in a session. Edges are undirected, which prevents sequential information from being modeled in the graph. Each edge is weighted by $\mathbf{e}_{ij}$ which denotes the normalized number of co-occurrences (in all sessions) of the two items it connects. The learning of item-item transitions is solely the task of the underlying sequential model and the item-item graph embedding only incorporates the similarity to other items. We strictly rely on inductive representations for new items to obviate any data leakage into the embeddings.

The item catalog can contain millions of unique items which strongly impacts the scalability of the generated item-item graph. Therefore, we apply Bootstrapped Graph Latents (BGRL)~\cite{thakoor2021bootstrapped} that can be scaled up to graphs with hundreds of millions of nodes and reduce the memory requirement substantially by enriching the original graph with simple augmentations to produce two different, but semantically similar views. Two encoders then generate online and target embeddings. The online embedding is used as input to a predictor which forms a prediction for the target embedding. The cosine similarity of the predictor output and the generated target embedding by the encoder is the final objective. Our graph embedding in GCNext is based on the BGRL learning method with our custom encoder architecture. We use the attentional convolution as introduced in~\cite{velickovic2018graph} and optimized in~\cite{brody2021attentive} as \textit{GATv2}. The node-wise formulation of this graph operation is defined as:
\begin{equation}
    \mathbf{h}_i^{'} = \sigma \left(\sum_{j \in \mathcal{N}_i} \alpha_{ij} \cdot \mathbf{W}\mathbf{h}_j\right),
\end{equation}
where $\mathbf{h}$ denotes the node features, $\mathbf{W}$ the linear transformation’s weight matrix and the average over the neighbor nodes features is weighted by the normalized attention weights $\alpha_{ij}$. The computation of the normalized attention coefficients with softmax including edge weights can be seen in:
\begin{equation}
    \alpha_{ij} =
    \frac{
    \exp\left(\mathbf{a}^{\top}\mathrm{LeakyReLU}\left(\mathbf{W} \cdot
    [\mathbf{h}_i \, \Vert \, \mathbf{h}_j \, \Vert \, \mathbf{e}_{ij}]
    \right)\right)}
    {\sum_{k \in \mathcal{N}_i}
    \exp\left(\mathbf{a}^{\top}\mathrm{LeakyReLU}\left(\mathbf{W} \cdot 
    [\mathbf{h}_i \, \Vert \, \mathbf{h}_k \, \Vert \, \mathbf{e}_{ik}]
    \right)\right)},
\end{equation}
where $\mathbf{W}$ is a learnable weight matrix, $\mathbf{h}$ again denotes the corresponding node features, $\mathbf{e}_{ij}$ is the edge weight from node $n_{i}$ to neighboring node $n_{j}$ and $\Vert$ denotes vector concatenation.

Along the lines of other encoder architectures~\cite{velickovic2019deep,thakoor2021bootstrapped}, we stack multiple graph convolutional layers using skip connections and PReLU~\cite{he2015delving} as an activation function. The second layer of the encoder performs the computation:
\begin{equation}
    \label{eq:skip-encoder}
    \mathbf{H_2} = \sigma(\mathrm{GATv2Conv}(\mathbf{H_1} + \mathbf{X}\mathbf{W}_{skip}, \mathbf{A})),
\end{equation}
where $\mathbf{H_1}$ is the output of the previous layer, $\mathbf{X}$ are the input node features, $\mathbf{W}_{skip}$ is a learnable projection matrix for the skip connection, and $\mathbf{A}$ denotes the adjacency matrix.

Since large graphs, \ie our proposed item co-occurrence graph, generally do not fit into GPU memory, we rely on a sub-graph sampling approach based on neighborhood batch sampling in~\cite{hamilton2017inductive}, subsampling a fine-tuned, fixed-sized neighborhood per node.

\subsection{Extension of Sequential Models}
In sequential neural network models, next items are predicted by multiplying the candidate item embeddings with the learned session representation and applying the softmax operation to obtain the corresponding item probabilities. To combine the advantage of graph-based item embeddings and current state-of-the-art models in session-based recommendation, we propose the following approach: Instead of initializing the sequential model's item embedding table with weights based on sophisticated initialization methods (for instance, the widely used Xavier initialization~\cite{glorot2010understanding}), we directly adopt the graph-generated item embeddings. Thereby, GCNext improves the learning process of the underlying sequential model as these embeddings capture topological knowledge about the item-item relations and contain additional item feature information.

We train the graph-based item embeddings in an unsupervised manner. Compared to end-to-end embedding learning, this has two advantages: First, GCNext is a modular method that can easily be applied to different sequential models without altering their architectures. Second, splitting up the training process into two stages supports the usage in production, since the training of sequential models with the pre-trained item embeddings converges faster.

In contrast to neural network models, nearest--neighbor methods do not make use of an item embedding table; they find similar sessions based on the input sequence to predict the next candidate item. To use GCNext in nearest--neighbor approaches, we propose using item graph embeddings to find similar session neighbors. To integrate item graph embeddings in nearest--neighbor methods, we compute the similarity of sessions based on the cosine distance of the embedding of each item in the input session to every item in the candidate session. A threshold value on the distance is adapted to find similar embeddings. Candidate sessions are then scored by the corresponding position mapping of the similar items based on the nearest--neighbor method or again by their cosine similarity; as defined by the $\mathrm{r}$-score in Equation~\ref{eq:score}~\cite{ludewig2021empirical}.
\begin{equation}\label{eq:score}
    \mathrm{r} ( S^{(\mathrm{i})}, S^{(\mathrm{c})} ) = \frac{ | T_{\mathrm{i,c}} | }{ \sqrt{ |S^{(\mathrm{i})}| }\sqrt{ |S^{(\mathrm{c})}| } },
\end{equation}
where $S^{(\mathrm{i})}$ and  $S^{(\mathrm{c})}$ refer to the sets of items in input and candidate session, respectively, and $T_{\mathrm{i,c}}$ represents the set of (pairs of) items in the input and candidate session with embedding similarity below the threshold.

\section{Experimental Setup}

\subsection{Datasets and Preprocessing}
To evaluate GCNext, we conduct experiments on three widely used datasets with different characteristics from the e-commerce and music domains. The \textbf{Diginetica}\footnote{\url{https://cikm2016.cs.iupui.edu/cikm-cup/}} dataset (CIKM Cup 2016) provides different item features; we use the category and price of each item as features (side information)~\cite{li2017neural,liu2018stamp}. The \textbf{Tmall}\footnote{\url{https://tianchi.aliyun.com/dataset/dataDetail?dataId=42}} dataset as part of the IJCAI-15 competition contains users' shopping logs along with the category, brand, and seller as additional item features~\cite{tang2018caser,zhang2019feature}. Furthermore, we evaluate GCNext on the \textbf{Music4All+} dataset, a version of the Music4All\footnote{\url{https://sites.google.com/view/contact4music4all}} dataset~\cite{santana2020music4all} with 11 item features. We enrich this dataset with i-vectors~\cite{dehak2010front} of dimensionality 100 based on the 13-dimensional Mel-Frequency Cepstral Coefficients of the songs and a Gaussian Mixture Model with 256 components~\cite{eghbal2015vectors}, using the \texttt{kaldi} toolkit~\cite{povey2011kaldi}.
Similar to~\cite{santana2020music4all}, we consider listening events to belong to the same session if there are no gaps of more than 30 minutes between them.


Following previous works~\cite{liu2018stamp,wu2019session}, we filter out items appearing less than 5 times and ignore sessions consisting of a single interaction. Additionally, training sequences are generated by splitting the input sequence into smaller sub-sequences. Consider, for example, the input sequence $s = [i_1, i_2,...,i_n]$, then the generated sequences and corresponding next items are $([i_1],i_2), ([i_1, i_2], i_3), ...,$ $([i_1,i_2,...i_{n-1}], i_n)$. The maximum sequence length is set to 50. Each dataset is sorted by its timestamps and temporally split into training, validation, and test set ($80\%$, $10\%$, and $10\%$). Table~\ref{tab:datasets} provides an overview of the datasets.

\begin{table}[h!]
    \caption{Dataset Statistics: Number of items, features, sessions, and average session length.}
    \label{tab:datasets}
    \scalebox{1.0}{
    \begin{tabular}{lrrrr}
    \toprule
    \textbf{Dataset}    & \textbf{Items} & \textbf{Feat.} & \textbf{Sessions} &\textbf{Avg. Length} \\ \midrule
    Diginetica          & 44,527        & 2             & 205,698      & 4.85        \\
    Tmall                & 97,259        & 3             & 188,113      & 8.11       \\
    Music4All+          &  80,471       & 12            & 601,858      & 7.70      \\
    \bottomrule
    \end{tabular}
    }
\end{table}


\subsection{Base Algorithms and Implementation}
One of the main advantages of GCNext is that it can be added to sequential models in a plug-in fashion to boost their performance. Therefore, we evaluate the following base algorithms for session-based recommendation with and without adding GCNext: SKNN~\cite{jannach2017recurrent}, STAN~\cite{garg2019sequence}, V--SKNN~\cite{ludewig2018effective}, and VSTAN~\cite{ludewig2021empirical} as representative nearest--neighbor methods;  
GRU4Rec+~\cite{tan2016improved}, Caser~\cite{tang2018caser}, NARM~\cite{li2017neural}, STAMP~\cite{liu2018stamp}, and SASRec~\cite{kang2018self} as state-of-the-art neural network models. 
In addition, we compare GCNext to 
current graph-based approaches (SR--GNN~\cite{wu2019session}, GCSAN~,\cite{xu2019graph} and LightGCN~\cite{he2020lightgcn}), and models including additional item features: GRU4RecF~\cite{hidasi2016parallel} and FDSA~\cite{zhang2019feature}. We additionally implement the graph-based approaches to incorporate the original item features in their initial embedding tables with naive sum-pooling, which are referred to as SR--GNNF, GCSANF, and LightGCNF.

All base models use the implementation in \textit{RecBole}~\cite{recbole} for neural network methods and \textit{session-rec}~\cite{ludewig2021empirical} for nearest-neighbor methods. For the BGRL implementation, we rely on the code given in~\cite{thakoor2021bootstrapped}. We use AdamW~\cite{loshchilov2017decoupled} to optimize item graph embeddings and Adam~\cite{kingma2014adam} in the sequential model training. The embedding size is fixed to~128 for comparability and all models performed best with this configuration according to preliminary experiments. We conduct hyperparameter optimization via grid search including the learning rate, number of layers and heads, layer sizes, and dropout rates for augmentation. Each experiment is repeated five times and the average results are reported. We provide our implementation on Github\footnote{\url{https://github.com/dbis-uibk/gcnext}}.


We adopt two widely used evaluation metrics to assess the quality of the recommendation lists: \emph{HR@k} (Hit Rate) and \emph{MRR@k} (Mean Reciprocal Rank). Similar to previous works~\cite{liu2018stamp,ren2019repeatnet,de2021transformers4rec}, each metric is computed with $k$ set to 10 and 20.

\section{Results and Analysis}
Table~\ref{tab:results} presents the experimental results of all base methods and their performance when extended with the proposed GCNext approach (column GCNext). When integrating GCNext into the nearest--neighbor methods (top part of Table~\ref{tab:results}), we observe a significant performance increase for some base models, and no significant decrease for any of the base models. In particular, V-SKNN's performance improves by $0.88\%$ to $6.33\%$ over all datasets on the \emph{HR@10} score. Although the metrics show that nearest--neighbor methods are able to keep up with certain neural network approaches, they lack in state-of-the art performance across all three datasets.

The effectiveness of our approach is distinctly indicated by the results of neural network approaches in session-based recommendation. Compared to all neural network methods, Caser, which uses convolution-based sequence embeddings, and STAMP, a complete attention-based method, benefit the most by incorporating GCNext across all three datasets. On the Diginetica dataset GCNext combined with SASRec, a transformer-based model, achieves the highest score on each metric throughout. This effect can also be seen with the Tmall dataset, where SASRec with GCNext increases the state-of-the-art \emph{HR@10} score by $1.12\%$ and significantly outperforms feature-based methods such as GRU4RecF and FDSA. STAMP extended with GCNext even achieves an increase in performance of $12.79\%$ on the \emph{MRR@20}. Interestingly, this effect becomes less impactful for the Music4All+ dataset which provides a large set of additional item features. On this dataset the graph-based self-attention network GCSAN extended with item features achieves the highest \emph{MRR} scores. Nonetheless, the evaluation shows significantly improved scores for each of the neural network-based models on the Music4all+ dataset. 

It is also noteworthy that neural network models trained with pre-trained item graph embeddings converge faster than their corresponding standard initialized counterparts. We assume this is due to the enriched information contained in the item embeddings.

\begin{table*}[ht!]
    \caption{Model performances on all datasets; column GCNext indicates the use of GCNext. Significant improvements over the underlying sequential models (paired \textit{t}-test, $p < .05$) are marked with $^\dagger$; best results in bold, second-best results underlined.}
    \label{tab:results}
    \scalebox{0.82}{
    \begin{tabular}{l|l|SSSS|SSSS|SSSS}
    \toprule
    \multirow{4}{*}{Model} &
    \multirow{4}{*}{\rotatebox[origin=l]{90}{GCNext}} &
      \multicolumn{4}{c}{Diginetica} &
      \multicolumn{4}{c}{Tmall} &
      \multicolumn{4}{c}{Music4All+}\\
      & & \multicolumn{2}{c}{MRR} & \multicolumn{2}{c}{HR} & \multicolumn{2}{c}{MRR} & \multicolumn{2}{c}{HR} & \multicolumn{2}{c}{MRR} & \multicolumn{2}{c}{HR}\\
      & & {@10} & {@20} & {@10} & {@20} & {@10} & {@20} & {@10} & {@20} & {@10} & {@20}  & {@10} & {@20}\\
      \midrule
      \multicolumn{14}{c}{\emph{Nearest--neighbor Methods}}\\
      \midrule
        SKNN & \xmark & 17.98 & 18.62 & 36.67 & 45.94 & 22.13 & 22.41 & 32.32 & 36.30 & 19.49 & 19.89 & 42.83 & 48.45 \\
        STAN & \xmark & 15.43 & 16.12 & 28.95 & 38.97 & 20.13 & 20.41 & 26.15 & 30.20 & 18.48 & 18.85 & 37.56 & 42.78 \\
        V-SKNN & \xmark & 17.99 & 18.63 & 36.59 & 45.84 & 21.84 & 22.09 & 30.95 & 34.55 & 19.58 & 19.99 & 43.06 & 48.68 \\
        VSTAN & \xmark & 15.27 & 15.97 & 28.64 & 38.73 & 20.04 & 20.32 & 26.01 & 30.04 & 18.41 & 18.77 & 36.99 & 42.18 \\
        \bottomrule
        SKNN & \cmark & 18.07$^\dagger$ & 18.71$^\dagger$ & 37.11$^\dagger$ & 46.40$^\dagger$ & 22.14 & 22.40 & 32.32 & 36.31 & 19.43 & 19.85 & 43.10$^\dagger$ & 48.90$^\dagger$\\
        STAN & \cmark & 15.43 & 16.12 & 28.95 & 38.97 & 20.14 & 20.42 & 26.16 & 30.21 & 18.49 & 18.85 & 37.60$^\dagger$ & 42.80$^\dagger$\\
        V-SKNN & \cmark & 18.18$^\dagger$ & 18.83$^\dagger$ & 37.23$^\dagger$ & 46.47$^\dagger$ & 22.38$^\dagger$ & 22.67$^\dagger$ & 32.91$^\dagger$ & 36.95$^\dagger$ & 19.66$^\dagger$ & 20.07$^\dagger$ & 43.44$^\dagger$ & 49.20$^\dagger$\\
        VSTAN & \cmark & 15.27 & 15.97 & 28.64 & 38.73 & 20.04 & 20.32 & 26.01 & 30.05 & 18.56$^\dagger$ & 18.92$^\dagger$ & 37.41$^\dagger$ & 42.57$^\dagger$\\
        \midrule        \multicolumn{14}{c}{\emph{Neural Network Methods}}\\
        \midrule        
        GRU4Rec+ & \xmark & 17.09 & 17.94 & 38.19 & 50.45 & 28.62 & 29.00 & 45.71 & 51.02 & 25.18 & 25.62 & 41.35 & 47.53\\
        Caser & \xmark & 14.29 & 14.82 & 26.54 & 34.20 & 24.30 & 24.54 & 35.61 & 39.04 & 19.26 & 19.62 & 31.51 & 36.63\\
        STAMP & \xmark & 16.38 & 17.17 & 35.62 & 47.04 & 21.63 & 21.88 & 31.62 & 35.21 & 28.12 & 28.45 & 41.96 & 46.60\\
        NARM & \xmark & 17.35 & 18.18 & 38.54 & 50.57 & 28.11 & 28.46 & 44.58 & 49.71 & 28.82 & 29.20 & 42.64 & 48.00\\
        SASRec & \xmark & \underline{19.88} & \underline{20.73} & \underline{43.09} & \underline{55.24} & \underline{29.46} & \underline{29.84} & \textbf{47.72} & \underline{53.15} & 28.83 & 29.23 & \underline{45.25} & \underline{51.01}\\
        \bottomrule
        GRU4Rec+ & \cmark & 17.43$^\dagger$ & 18.27$^\dagger$ & 38.59$^\dagger$ & 50.78$^\dagger$ & 28.74$^\dagger$ & 29.12$^\dagger$ & 46.11$^\dagger$ & 51.60$^\dagger$ & 28.78$^\dagger$ & 29.18$^\dagger$ & 43.33$^\dagger$ & 48.99$^\dagger$\\
        Caser & \cmark & 15.63$^\dagger$ & 16.28$^\dagger$ & 30.60$^\dagger$ & 39.99$^\dagger$ & 26.33$^\dagger$ & 26.68$^\dagger$ & 41.15$^\dagger$ & 46.12$^\dagger$ & 20.56$^\dagger$ & 20.92$^\dagger$ & 33.35$^\dagger$ & 38.54$^\dagger$\\
        STAMP & \cmark & 17.43$^\dagger$ & 18.23$^\dagger$ & 37.50$^\dagger$ & 49.08$^\dagger$ & 24.33$^\dagger$ & 24.68$^\dagger$ & 37.75$^\dagger$ & 42.82$^\dagger$ & 28.60$^\dagger$ & 28.95$^\dagger$ & 42.81$^\dagger$ & 47.87$^\dagger$\\
        NARM & \cmark & 17.89$^\dagger$ & 18.73$^\dagger$ & 39.51$^\dagger$ & 51.73$^\dagger$ & 28.99$^\dagger$ & 29.39$^\dagger$ & 46.48$^\dagger$ & 52.19$^\dagger$ & 28.89$^\dagger$ & 29.28$^\dagger$ & 43.11$^\dagger$ & 48.67$^\dagger$\\
        SASRec & \cmark & \textbf{19.97}$^\dagger$ & \textbf{20.80}$^\dagger$ & \textbf{43.10} & \textbf{55.41}$^\dagger$ & \textbf{29.51}$^\dagger$ & \textbf{29.93}$^\dagger$ & \underline{47.70} & \textbf{53.75}$^\dagger$ & 29.84$^\dagger$ & 30.15$^\dagger$ & \textbf{45.43}$^\dagger$ & \textbf{51.13}$^\dagger$\\
        \midrule        \multicolumn{14}{c}{\emph{Feature \& Graph-based Methods}}\\
        \midrule        
        GRU4RecF & \xmark & 16.04 & 16.91 & 36.35 & 48.77 & 25.25 & 25.66 & 42.04 & 47.97 & 28.67 & 29.04 & 42.52 & 47.80\\
        FDSA & \xmark & 18.92 & 19.79 & 41.29 & 53.73 & 28.76 & 29.15 & 46.48 & 52.01 & \underline{30.04} & \underline{30.43} & 45.20 & 50.66\\
        SR-GNN & \xmark & 17.75 & 18.58 & 39.18 & 51.23 & 27.47 & 27.84 & 44.67 & 49.89 & 28.90 & 29.27 & 42.62 & 47.91\\
        SR-GNNF & \xmark & 17.49 & 18.33 & 38.45 & 50.56 & 25.55 & 25.98 & 41.52 & 47.70 & 28.72 & 29.08 & 41.98 & 47.14\\
        GCSAN & \xmark & 19.20 & 20.03 & 41.00 & 53.01 & 29.01 & 29.41 & 47.37 & 53.04 & 29.68 & 30.05 & 43.79 & 49.09\\
        GCSANF & \xmark & 17.21 & 18.04 & 37.60 & 49.57 & 25.16 & 25.58 & 40.60 & 46.67 & \textbf{30.13} & \textbf{30.49} & 43.74 & 48.84 \\
        LightGCN & \xmark & 15.90 & 16.67 & 35.11 & 46.29 & 26.03 & 26.47 & 45.92 & 52.16 & 8.43 & 9.01 & 21.98 & 30.18\\
        LightGCNF & \xmark & 15.90 & 16.66 & 34.96 & 45.92 & 25.80 & 26.27 & 45.49 & 52.25 & 10.95 & 11.59 & 28.49 & 37.45\\
        \bottomrule
      \end{tabular}}
\end{table*}

\section{Conclusion and Future Work}
We proposed GCNext, an extension to sequential recommendation models based on GCNs. In the first phase, we generate an item co-occurrence graph with nodes/items enriched with item descriptors to learn its node representations in an unsupervised manner. In the second phase, we use the learned item embedding weights to initialize the item embedding table of the underlying sequential base model. Our experimental results on three different datasets show the effectiveness of GCNext. 

For future work, we plan to further investigate the potential improvements of graph-based item embeddings in cold-start scenarios for session-based recommendation. Additionally, future experiments will compare our approach with different graph embeddings methods and investigate the diverging impact on baseline models.

\begin{acks}
This research was funded in whole, or in part, by the Austrian Science Fund (FWF) [P33526]. For the purpose of open access, the author has applied a CC BY public copyright licence to any Author Accepted Manuscript version arising from this submission.
\end{acks}

\bibliographystyle{ACM-Reference-Format}
\bibliography{bibliography}

\end{document}